\documentclass[aps,pra,twocolumn,floatfix,nofootinbib,superscriptaddress,graphics,longbibliography]{revtex4-2}
\usepackage{bm,graphicx,mathrsfs,amsmath,amssymb,mathtools,makecell,amssymb,bbm, amsthm,dsfont,color,times,txfonts,nicefrac,framed,float,enumitem,tikz,physics}
\usepackage{algorithm}
\usepackage{algpseudocode}
\usepackage{amsfonts}
\usepackage[normalem]{ulem}
\usepackage[colorlinks=true,linkcolor=equationcolor,citecolor=teal]{hyperref}
\usepackage{tcolorbox}\tcbset{before skip=10pt,toptitle=1mm,bottomtitle=1mm,fonttitle=\bfseries}
\tcbuselibrary{theorems}
\tcbuselibrary{breakable}
\newtcolorbox[auto counter]{pabox}[2][]{%
	colback=white,colframe=black!10,fonttitle=\bfseries,coltitle=black,breakable,title=#2,#1}

\hypersetup{urlcolor=teal}
\definecolor{equationcolor}{RGB}{222,94,100}

\newcommand{\ketbraa}[2]{\ensuremath{\left|#1\right\rangle\!\!\left\langle#2\right|}}

\newcommand{\ms}[1]{\textsf{#1}}
\newcommand{\ot}{\otimes}

\newcommand{\cat}{\chi}
\begin{document}

\title{Quantum catalysis in cavity quantum electrodynamics}

\author{A. de Oliveira Junior}
\affiliation{Faculty of Physics, Astronomy and Applied Computer Science, Jagiellonian University, 30-348 Kraków, Poland.}
\author{Mart\'i Perarnau-Llobet}
 \affiliation{Department of Applied Physics, University of Geneva, 1211 Geneva, Switzerland}
 \author{Nicolas Brunner}
  \affiliation{Department of Applied Physics, University of Geneva, 1211 Geneva, Switzerland}
\author{Patryk Lipka-Bartosik}
\affiliation{Department of Applied Physics, University of Geneva, 1211 Geneva, Switzerland}

\date{\today}

% ------------------------------------------------
%   ABSTRACT
% ------------------------------------------------

\begin{abstract}

Catalysis plays a key role in many scientific areas, most notably in chemistry and biology. Here we present a catalytic process in a paradigmatic quantum optics setup, namely the Jaynes-Cummings model, where an atom interacts with an optical cavity. The atom plays the role of the catalyst, and allows for the deterministic generation of non-classical light in the cavity. Considering a cavity prepared in a ``classical'' coherent state, and choosing appropriately the atomic state and the interaction time, we obtain an evolution with the following properties. First, the state of the cavity has been modified, and now features non-classicality, as witnessed by sub-Poissonian statistics or Wigner negativity. Second, the process is catalytic, in the sense that the atom is deterministically returned to its initial state exactly, and can be re-used multiple times. What is more, we also show that our findings are robust under dissipation and can be applied to scenarios featuring cavity loss and atomic decay. Finally, 
we investigate the mechanism of this catalytic process, in particular highlighting the key role of correlations and quantum coherence. 

\end{abstract}

\maketitle

\section{Introduction}
The effect of catalysis involves using an auxiliary system (a catalyst) to enable a process that would either not occur spontaneously or would occur very slowly. Catalysis manifests across a variety of fields (see e.g.~\cite{chorkendorff2017concepts}), including biological processes activated by enzymes, the speed-up of chemical reactions, and the synthesis of nanomaterials. 

More recently, the phenomenon of catalysis has also become relevant in the context of quantum information; see recent reviews \cite{Datta2022,Patrykreview}. First examples focused on entanglement manipulation~\cite{Jonathan_1999,vanDam2003,turgut2007catalytic,daftuar2001mathematical,sun2005existence,feng2005catalyst,PhysRevLett.127.080502,Kondra_2021,Datta2022}, and then spread to quantum thermodynamics~\cite{Brand_o_2015,Ng_2015,Wilming2017,Mueller2018,Lipka-Bartosik2021,shiraishi2021quantum,Gallego2016,boes2019bypassing,Henao_2021,Henao2022,son2022catalysis,son2023hierarchy,Lipka_Bartosik_2023,czartowski2023thermal}, coherence theory~\cite{Aberg2014,Vaccaro2018,lostaglio2019coherence,takagi2022correlation,char2023catalytic,van2023covariant} and other areas~\cite{Marvian2019,wilming2021entropy,Wilming2022correlationsin,rubboli2022fundamental,lie2021catalytic,boes2019neumann}. These results are typically formulated within the framework of quantum resource theories~\cite{chitambarGour}. This abstract approach is particularly useful for characterizing the fundamental limits of manipulating quantum resources, including scenarios involving catalytic systems of arbitrary complexity. 

\begin{figure}[t]
    \centering
    \vspace{1.1cm}
    \includegraphics{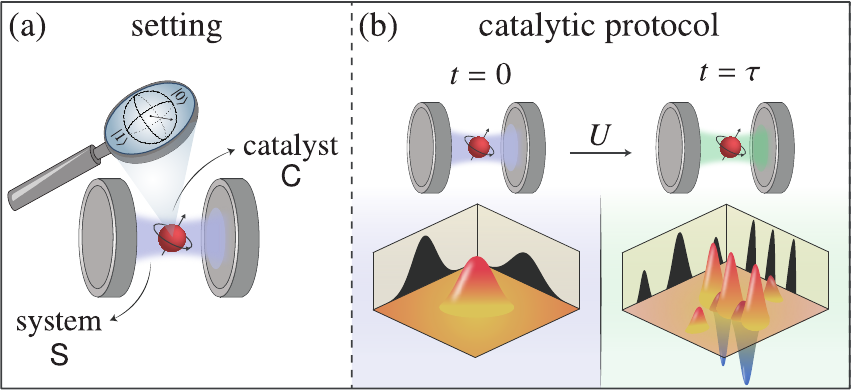}
    \caption{\textbf{Quantum catalysis in the Jaynes-Cummings model.} (a) An atom (the catalyst $\ms C$) interacts with a single-mode optical cavity (the system $\ms S$), initially prepared in a ``classical'' coherent state. (b) We consider the evolution $U$ over a well-chosen time interval (from $t=0$ to $t=\tau$) such that (i) the final state of the cavity is non-classical, and (ii) the atom is returned to its intial state exactly. Hence, non-classicality has been generated via catalysis.}
    \label{scheme}
\end{figure}

An interesting direction is whether quantum catalysis is also relevant and useful in a more practical context, potentially even in experiments. Here we investigate quantum catalysis in a paradigmatic setup of quantum optics, namely the Jaynes-Cummings model~\cite{Jaynes1963,Larson_2021,Greentree_2013}, where a two-level atom interacts with a single-mode optical cavity. We uncover a catalytic process enabling the generation of a non-classical state of light in the cavity, using the atom as a catalyst. Specifically, we consider the cavity to be initially prepared in a ``classical'' coherent state, and uncorrelated to the atom. By carefully setting the initial state of the atom and the interaction time, we obtain a final state such that (i) the atom is back in its initial state exactly, and (ii) the state of the cavity is now non-classical, i.e. featuring Wigner negativity or sub-Poisonian statistics. Hence, non-classicalilty of the cavity has been generated without perturbing the state of the atom (see Fig.~\ref{scheme}). The process is catalytic and the atom could be re-used, for example by coupling it to another cavity. 

We investigate the mechanism of this catalytic process, and identify two crucial ingredients. First, the final state of the atom and cavity must feature correlations. Second, the evolution of the state of the catalyst must involve quantum coherence (i.e. superpositions of the energy basis states). The latter is an interesting aspect, as typical instances of quantum catalysis in resource theories involve only diagonal states (i.e. without coherence), so that they can be understood as a stochastic process involving the probability distributions of the system and catalysis. Instead here, the system and catalyst experience a genuinely quantum evolution. This is a novel instance of the effect of \emph{coherent quantum catalysis}, recently investigated in quantum thermodynamics~\cite{Lipka_Bartosik_2023}.

Before proceeding, it is worth discussing previous works in quantum optics that relate to the concept of catalysis. Notably, the pioneering proposal for quantum computing in ion traps \cite{Cirac1995} (see also~\cite{Phoenix1993,Hagley1997}) considers two spin qubits that become entangled via an interaction with a cavity that can be considered catalytic~\cite{Patrykreview}. Another relevant direction is that of ``multi-photon catalysis'' (see e.g.~\cite{Lvovsky2002,Bartley2012,hu2016multiphoton,Hu2017}) which is a heralded catalytic process, where the catalyst is returned only with some probability. In contrast, our catalytic protocol is deterministic. 

% ------------------------------------------------
% SEC. I - CATALYSIS IN THE JAYNES-CUMMINGS MODEL
% ------------------------------------------------

\section{Catalysis in the Jaynes-Cummings model} \label{sec:framework}

Consider a setup comprising a system $(\ms{S})$ initially prepared in a state $\rho_{\ms S}$ and a catalyst $(\ms{C})$ in an initial state $\cat_{\ms C}$. The total system $\ms{SC}$ is assumed to be closed and evolves via an energy-conserving process for some time $\tau$. This evolution is represented by the unitary $U=\exp({-i H_{\ms{SC}} \tau})$, where $H_{\ms{SC}}$ denotes the Hamiltonian. Consequently, the final state of the total system is given by $\sigma_{\ms{SC}} := U(\rho_{\ms{S}} \ot \cat_{\ms{C}})U^{\dagger}$.

The evolution is said to be \emph{catalytic} when the catalyst is returned in exactly the same state as it was initially prepared. Formally, we demand that
\begin{equation}\label{Eq:catalytic-constrain}
    \sigma_{\ms C} := \Tr_{\ms{S}}[U(\rho_{\ms{S}} \ot \cat_{\ms{C}})U^{\dagger}] = \cat_{\ms C},
\end{equation}
which we refer to as the \emph{catalytic constraint}. Satisfying this constraint typically requires to carefully choose the initial states of the system $\rho_{\ms{S}}$ and the catalyst $\cat_{\ms{C}}$, as well as the interaction time $\tau$.

The main goal of a catalytic evolution is to induce an interesting local dynamics on the system $\ms{S}$, i.e.
\begin{align} \label{eq:local_evol_S}
    \rho_{\ms{S}} \rightarrow \sigma_{\ms{S}} := \Tr_{\ms{C}}[U(\rho_{\ms{S}} \ot \cat_{\ms{C}})U^{\dagger}],
\end{align}
while leaving the state of the catalyst unchanged. Notably, it is possible to induce an evolution on $\ms{S}$ [as in Eq.~\eqref{eq:local_evol_S}] that would not be possible without the presence of the catalyst. 

In this work, we discuss the phenomenon of catalysis in the Jaynes-Cummings (JC) model, describing the interaction between a single-mode optical cavity and a two-level atom~\cite{Jaynes1963} (see Fig.~\hyperref[scheme]{\ref{scheme}a}). We choose the cavity to represent the system $\ms S$, while the atom will play the role of the catalyst $\ms C$. The cavity is characterized by the bosonic annihilation operator $a$ with the photon number operator $n_{\ms{S}} := a^{\dagger}a$. The atom has energy levels $\ket{g}$ and $\ket{e}$ and its energy is captured by $\sigma_z = \ketbra{e}{e}-\ketbra{g}{g} $. We work in the resonant regime, where the atom and cavity have the same frequency $\omega$. The evolution is governed by the JC Hamiltonian, which in the rotating-wave approximation reads 
\begin{align}\label{Eq:Total_Hamiltonian}
    H_{\ms{SC}} = \omega a^{\dagger} a + \frac{\omega}{2} \sigma_z +   g \left(\sigma_+ a + \sigma_- a^{\dagger} \right),
\end{align}
where $g$ is the coupling constant and \mbox{$\sigma_{+} = \ketbra{e}{g}$} and \mbox{$\sigma_- = \ketbra{g}{e}$}. Note that, as we focus on the resonant regime, the evolution specified by Eq. (\ref{Eq:Total_Hamiltonian}) is energy-preserving. 

Our goal is to catalytically generate non-classical light in the cavity starting from a (classical) coherent state, i.e.
\begin{equation} \label{eq:coherent-state}
    \ket{\alpha} = e^{-|\alpha|^2/2}\sum_{n = 0}^{\infty} \frac{\alpha^n}{\sqrt{n !}} \ket{n}.
\end{equation}
This state has Poissonian statistics with the mean number of photons $\langle n_{\ms{S}}\rangle_{\dyad{\alpha}} = |\alpha|^2$. 

\begin{figure}[t]
    \centering
    \includegraphics{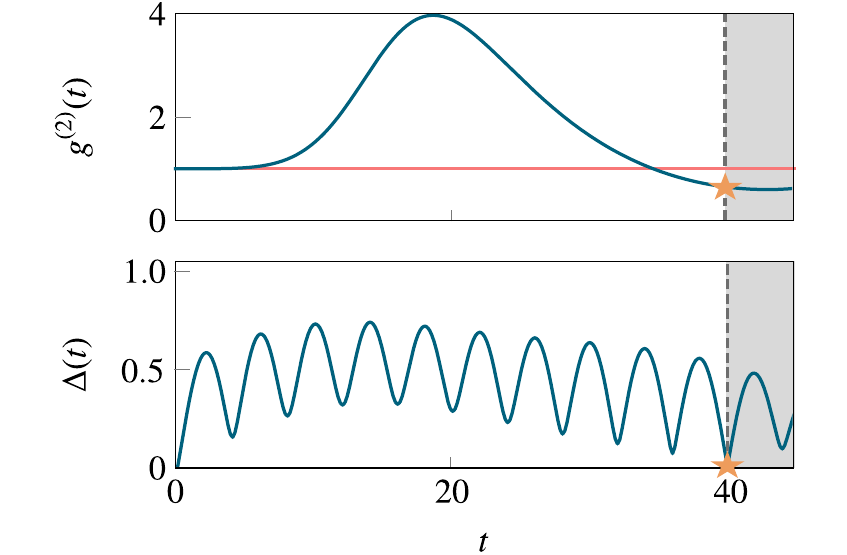}
    \caption{\textbf{First illustrative example.} Catalytic process for generating non-classicality in the cavity, as captured by the second-order auto-correlation function. Its time evolution, $g^{(2)}(t) := g^{(2)}(\sigma_{\ms{S}}(t))$, is shown in the top panel, while the bottom panel shows the modification of the atomic state, via the distance $\Delta(t) := \norm{{\cat_{\ms{C}}-\sigma_{\ms{C}}(t)}}_1$ with respect to the initial state. The orange stars indicate the final time ($\tau \approx 40$) for which the evolution is catalytic. The atom returns to its intial state ($\Delta(\tau)=0$), while non-classicality has been activated $g^{(2)}(\tau) \approx 0.5< g^{(2)}(0)=1 $. The value of the parameters are: $\alpha = 1/\sqrt{2}$, $\omega = 2\pi$, $g = \pi$.}
    \label{fig:2}
\end{figure}

In this work, we will derive a general characterisation of catalytic processes for the JC model. In particular,  we
\begin{enumerate}
 \item  Derive  a necessary and sufficient condition for the initial state of the atom $\cat_{\ms{C}}$ and an interaction time $\tau$ so that  the catalytic constraint \eqref{Eq:catalytic-constrain} is satisfied,  \item Give an analytic expression for the final state of the cavity after a catalytic process, 
\item  Develop an intuitive understanding of the mechanism behind catalysis in terms of the higher moments of observables, and finally 
\item  Show that catalytic evolutions can generate non-classicality in the state of the cavity, even in the presence of dissipation. 
 \end{enumerate}
 Before presenting these general results in Secs~\ref{sec:meachanism}, \ref{sec:how} and \ref{sec:catalysis-realistic-scenarios}, it is illustrative to discuss two specific examples of catalytic processes using complementary figures of merit to witness non-classicality.

\begin{figure*}
    \centering
    \includegraphics{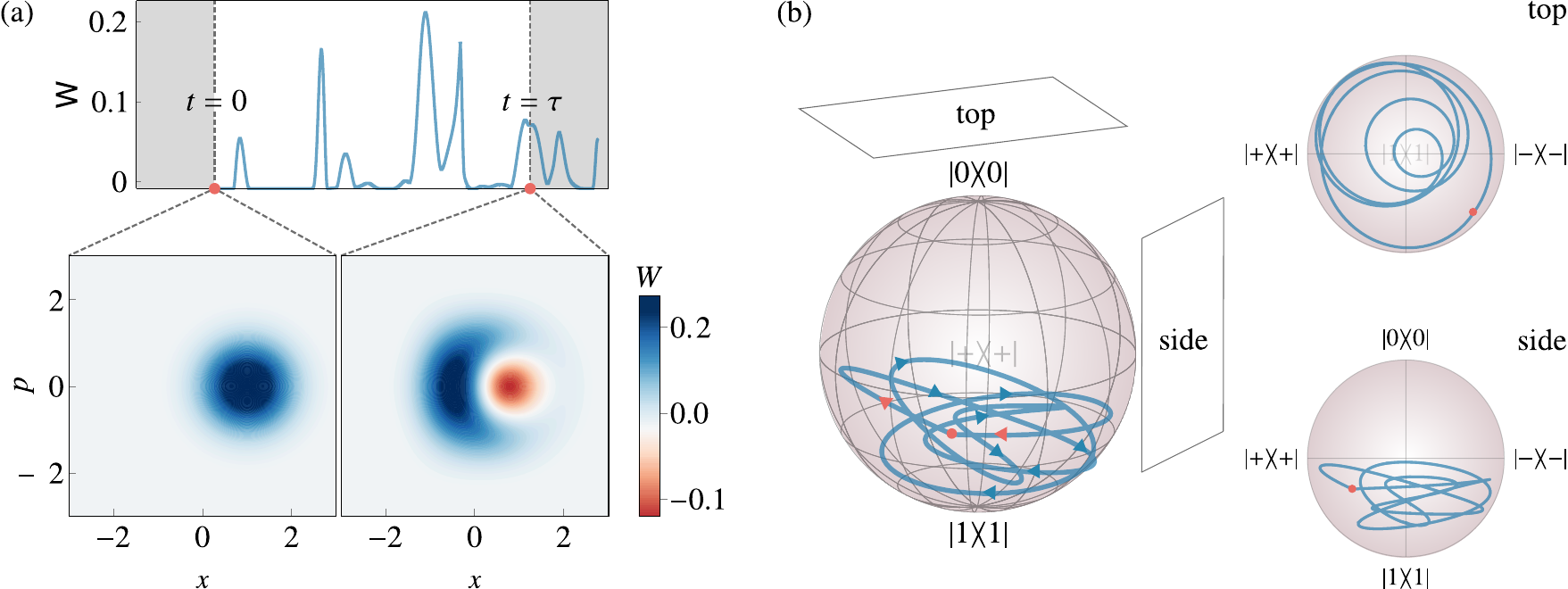}
    \caption{\textbf{Second illustrative example.} Catalytic generation of Wigner negativity. Panel (a) shows the time evolution of the Wigner logarithmic negativity $\mathsf W$, as well as the Wigner function at initial and final time $\tau$. Non-classicality is clearly generated, while the process is catalytic. Panel (b) shows the trajectory of the atomic state, i.e. the catalyst $\ms{C}$, on the Bloch sphere. In the trajectory, the state always remains inside the Bloch ball and never touches the surface. This is because states on the surface represent pure states, which cannot be used as catalysts. The initial and final state coincide (red dot). The values of the parameters are: $\alpha = 1/\sqrt{2}$, $\omega = 2\pi$ and $g = \pi$.}
    \label{F-Wigner-WLN-nav}
\end{figure*}

For our first example, we quantify non-classicality via the second-order auto correlation function of the final state of the cavity~\cite{PhysRev.130.2529} , i.e.
\begin{align}\label{eq:def_g2}
   g^{(2)}(\sigma_{\ms{S}}) = \frac{\langle n_{\ms{S}}^2 \rangle_{\sigma}  - \langle n_{\ms{S}} \rangle_{\sigma}}{\langle n_{\ms{S}} \rangle_{\sigma}^2},
\end{align}
where $\langle n_{\ms{S}}\rangle_{\rho} := \Tr[\rho_{\ms{S}} n_{\ms{S}}]$. Our goal is to obtain $g^{(2)}(\sigma_{\ms{S}}) <1$, which certifies non-classicality.

In Fig.~\ref{fig:2}, we show such a catalytic generation of non-classicality. The upper plot depicts the time evolution of $g^{(2)}$. Catalysis occurs at time $\tau = 40$ (represented by orange stars). Starting from the initial state of the cavity $\rho_{\ms{S}} = \dyad{\alpha}_{\ms{S}}$ (with $\alpha=1/\sqrt{2}$), which has $g^{(2)}(\rho_{\ms{S}})=1$ as any coherent state, we obtain a final state $\sigma_{\ms{S}}$ for which $g^{(2)}(\sigma_{\ms{S}}) \approx 0.5$, hence non-classical. In parallel, we monitor the time evolution of the atom by calculating the trace distance to its initial state, i.e. 
\begin{align}
\Delta(t) := \norm{{\cat_{\ms{C}}-\sigma_{\ms{C}}(t)}}_1.
\end{align}
At time $\tau \approx 40$, we find $\Delta(\tau)=0$, indicating that the atom has returned to its initial state as required for catalysis.

As a second example, we demonstrate how Wigner negativity can be catalytically generated. We focus on the Wigner logarithmic negativity~\cite{veitch2014resource, Albarelli_2018,PhysRevA.97.062337} (WLN), defined as
\begin{align} \label{Wigner Neg}
\mathsf{W} \left ( \rho \right) := \log \left( \int \! \mathrm{d} x \,\mathrm{d} p \, \left| W_\rho \left(x,p \right) \right| \right),
\end{align}
where $W_\rho \left(x,p \right)$ is the Wigner function
\begin{align}\label{eq:w_fun}
    W_{\rho}(x,p) = \frac{1}{\pi} \int e^{2i p x'} \langle x-x'|\rho|x+x' \rangle \, \text{d}x'.
\end{align}
Starting from a coherent state $\rho_{\ms{S}} = \dyad{\alpha}_{\ms{S}}$ which has a positive Wigner function, hence $\mathsf{W}(\rho_{\ms{S}})=0$, we aim to obtain a final state of the cavity $\sigma_{\ms{S}}$ with $\mathsf{W}(\sigma_{\ms{S}})>0$, therefore certifying its non-classicality. In Fig.~\ref{F-Wigner-WLN-nav}, we present an example of such an evolution. First, we plot WLN as a function of time $t$. At the final time $\tau = 5$, we obtain a state $\sigma_{\ms{S}}$ that has $\mathsf{W}(\sigma_{\ms{S}}) \approx 0.1$, and we plot its Wigner function. To verify the catalytic nature of the evolution, we display the evolution of the atomic state via its trajectory in the Bloch sphere. Crucially, the trajectory is closed, as the initial and final state of the atom exactly coincide (red dot). 
Note that the two examples presented here complementary. The first one shows the generation of non-classicality as witnessed by the $g^{(2)}$ function, while the final state $\sigma_{\ms{S}}$ has a positive Wigner function. In the second example, the final state $\sigma_{\ms{S}}$ has a negative Wigner function, even though $g^{(2)}(\sigma) > 1$.

% ------------------------------------------------
% SEC. II - MECHANISM OF CATALYSIS
% ------------------------------------------------

\section{Mechanism of catalysis} \label{sec:meachanism}

Here we provide an intuitive understanding of quantum catalysis by identifying a mechanism which allows for generating non-classicality. This allows us to characterize the catalytic regime analytically.

In the Jaynes-Cummings model, the energies of the cavity $\ms{S}$ and the atom $\ms{C}$ are specified by number operators $n_{\ms{S}}$ and $n_{\ms{C}}$ respectively. Hence, the total energy of both systems is proportional to the number of excitations, and described by a joint operator \mbox{$n_{\ms{SC}} := n_{\ms{S}} + n_{\ms{C}}$}. Since the JC evolution $U(t)$ that takes $\rho_{\ms{S}} \ot \cat_{\ms{C}}$ into $\sigma_{\ms{SC}}$ conserves the total energy, we have that $[U(t), n_{\ms{S}} + n_{\ms{C}}] = 0$ for all $t$. When the evolution is catalytic, all moments of $n_{\ms{C}}$ must remain unchanged, in particular $\langle n_{\ms{C}} \rangle_{\cat}$ = $\langle n_{\ms{C}} \rangle_{\sigma}$ and consequently  $\langle n_{\ms{S}} \rangle_{\rho} = \langle n_{\ms{S}} \rangle_{\sigma}$. Importantly,  this is not the case for higher moments of $n_{\ms{S}}$. Particularly, in Appendix~\ref{App:higher-moments-of-observables} we show that the second moment satisfies
\begin{align}\label{eq:cat_variance_formula}
    \langle n^2_{\ms{S}} \rangle_{\sigma} = \langle n^2_{\ms{S}} \rangle_{\rho} + 2\Bigg(\langle n_{\ms{S}}\rangle_{\sigma} \langle n_{\ms{C}}\rangle_{\sigma} - \langle n_{\ms{S}}  \ot n_{\ms{C}}\rangle_{\sigma}\Bigg),
\end{align}
Hence, the second moment in the final state of the cavity, $\langle n^2_{\ms{S}} \rangle_{\sigma}$, can become smaller (or larger) than the second moment in the initial state $\langle n^2_{\ms{S}} \rangle_{\rho}$. This means that using a catalyst allows for modifying the distribution of the local observable $n_{\ms{S}}$ of the system $\ms{S}$: while its average must remain the same, the higher moments can change. Importantly, this can only happen if the system becomes correlated with the catalyst, i.e. $\langle n^2_{\ms{S}} \rangle_{\sigma} \neq \langle n^2_{\ms{S}} \rangle_{\rho}$ only if $\sigma \neq \sigma_{\ms{S}} \ot \sigma_{\ms{C}}$ [as seen from Eq. \eqref{eq:cat_variance_formula}]. Thus, correlations are essential for observing quantum catalysis. Finally, the above analysis also applies beyond the JC model, see Appendix~\hyperref[Appsub:highermomentsofobservables]{B-1a} for a generalisation of Eq. (\ref{eq:cat_variance_formula}) to arbitrary observables and moments. This can help to discover and understand new instances of quantum catalysis. 

The above analysis will serve as a basis for the characterisation of the parameter regime leading to catalysis. In particular we derive a necessary and sufficient condition for satisfying the catalytic constraint \eqref{Eq:catalytic-constrain}. Moreover, we obtain an analytic expression for the auto-correlation function~$g^2$, which is based on the second moment $ \langle n^2_{\ms{S}} \rangle_{\sigma}$.

We consider an arbitrary initial state of the atom 
\begin{align}
    \cat_{\ms C}= q \ketbraa{g}{g}+ r\ketbraa{g}{e}+r^*\ketbraa{e}{g}+[1-q] \ketbraa{e}{e},
\end{align}
as well as a general initial state of the cavity $\rho_{\ms{S}} = \sum_{n,m}^{\infty} p_{n,m}\ketbra{n}{m}$ with $ p_n :=  p_{n,n}$. Combining Eq. \eqref{eq:cat_variance_formula} with the fact that $\mbox{$\langle n_{\ms{S}} \rangle_{\sigma} = \langle n_{\ms{S}} \rangle_{\rho}$}$, we get
\begin{align}\label{eq:g2_conserved}
    \!g^{(2)}(\sigma_{\ms{S}}) = g^{(2)}(\rho_{\ms{S}}) - \frac{2}{\langle n_{\ms{S}}\rangle_{\rho}^2} \!\left[\langle n_{\ms{S}} \ot n_{\ms{C}}\rangle_{\sigma} \!- (1-q) \langle n_{\ms{S}} \rangle_{\rho}\right]\!,\!
\end{align}
where
\begin{align} \label{eq:corr_term}
    \!\!\!\langle n_{\ms{S}} \ot n_{\ms{C}}\rangle_{\sigma} \!=\! \sum_{n = 0}^{\infty} n \left[(1 - q) p_{n} c_n^2 + y_n + q p_{n+1} s_n^2 \right],
\end{align}
with $s_n := \sin(gt\sqrt{n+1})$, $c_n := \cos(g t \sqrt{n+1})$, and $y_{n} := 2\operatorname{Im}\qty[r p_{n+1,n}]s_n c_n$. For details see Appendix~\hyperref[Appsub:set-catalytic-states]{A-3a}.

In order to satisfy the catalytic constraint, we obtain a set of equations for the components of the atomic state. Decomposing the diagonal term as \mbox{$q = q_{\text{inc}} +  q_{\text{coh}}$}, we get
 \begin{align}\label{Eq-population}
 q_{\text{inc}} = \frac{1}{Q} \sum_{n=0}^{\infty}p_{n} s_n^2, \qquad  
q_{\text{coh}} = \frac{1}{Q} \sum_{n=0}^{\infty}y_n,  
\end{align}
with $Q := \sum_{n=0}^{\infty}(p_{n}+p_{n+1})s_n^2$. Interestingly, $q_{\text{inc}}$ is specified by the occupations of $\rho_{\ms{S}}$, while $q_{\text{coh}}$ depends on its coherence in the Fock basis. Moreover, the off-diagonal term $r$ satisfies 
\begin{equation}\label{Eq-coherence}
    r = \frac{i (a_3 a^{*}_4+a^{*}_1a_4)}{|a_1|^2 - |a_3|^2}-\frac{i(a_3 a^{*}_2+a^{*}_1a_2)}{|a_1|^2 - |a_3|^2}q,
\end{equation}
with $a_i$ being auxiliary functions defined as
\begin{align}\label{eq:aterms}
\begin{split}
    a_1 &= \sum_{n=0}^{\infty} p_{n,n}c_{n-1}c_{n}- e^{-i\omega \tau}, \qquad
    a_3 = \sum_{n=0}^{\infty} p_{n,n+2} s_ns_{n+1},  \\
    a_2 &= \sum_{n=0}^{\infty}p_{n,n+1}s_n\qty[c_{n-1}+c_{n+1}], \hspace{10pt}
    a_4 = \sum_{n=0}^{\infty}p_{n,n+1}s_nc_{n+1}. 
\end{split}
\end{align}
For a detailed derivation of Eqs. (\ref{eq:aterms}) see Appendix~\hyperref[Appsub:set-catalytic-states]{A-1}.

Importantly, Eqs \eqref{Eq-population} and \eqref{Eq-coherence} are necessary and sufficient for ensuring the catalytic constraint of Eq. \eqref{Eq:catalytic-constrain}. Combined with Eq. \eqref{eq:g2_conserved}, they characterize analytically the catalytic generation of non-classicality in our first example of a catalytic process (see Fig.~\ref{fig:2}). For the second example, we verify the catalytic constraint using our analytical findings and compute Wigner functions numerically.

A relevant question one might have is whether the atom can be reused in a \emph{truly} catalytic way during the catalytic protocol. That is, whether the catalytic protocol can be used to prepare multiple non-classical states in different cavities, while the catalyst is prepared only once. In Appendix \ref{app:mcav} we discuss this more operational view of catalysis in detail. 

% ------------------------------------------------
% SEC. III - HOW GENERAL IS CATALYSIS?
% ------------------------------------------------

\section{How general is catalysis?} \label{sec:how}

\begin{figure}[t]
    \centering
    \includegraphics{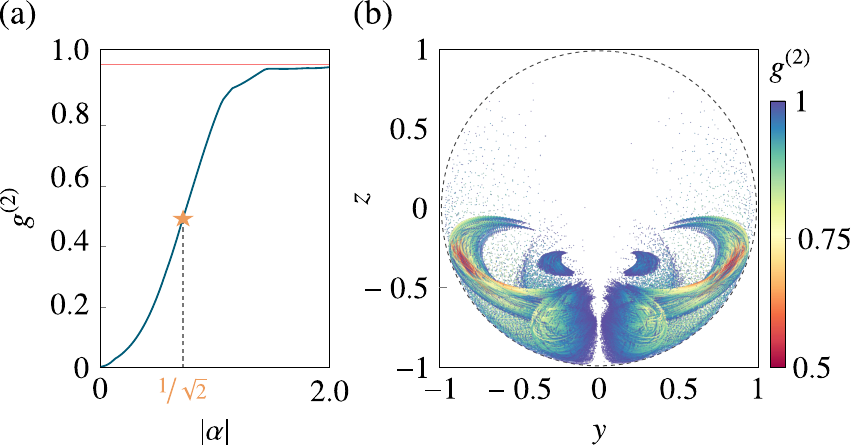}
    \caption{\textbf{Which states lead to catalysis?} Panel (a) shows the minimal value of $g^2$ obtained as a function of the amplitude $|\alpha|$ of the initial coherent state of the cavity $\rho_{\ms{S}} = \dyad{\alpha}_{\ms{S}}$. The orange star corresponds to our first illustrative example. Panel (b) displays the atomic states (in the $y-z$ plane of the Bloch sphere) that satisfy the catalytic constraint and generate non-classical states, for an intial coherent state $\alpha=1/\sqrt{2}$. The colour represents different values of $g^{(2)}<1$. We impose a limit on the interaction time $\tau\leq 100$ and take $10^6$ samples. In both panels, parameters are $\omega = 2\pi$, $g = \pi$.}
    \label{fig:1}
\end{figure}

An interesting problem is to understand how typical the effect of catalysis is. Here we discuss different aspects of this question. 

First, note that it is unclear a priori whether the catalytic constraint of Eq.~\eqref{Eq:catalytic-constrain} can be satisfied. Nonetheless, the quantum version of the Perron-Frobenius theorem guarantees that every quantum channel has at least one fixed point~\cite{fannes1992finitely}~\footnote{The existence of a catalytic state for an open dynamics can also be interpreted through this result.}. Now, for fixed $\rho_{\ms{S}}$ and $\tau$, the atom evolves according to an effective quantum channel $\cat_{\ms{C}} \rightarrow \Tr_{\ms{S}}[U(\tau)(\rho_{\ms{S}}\ot \cat_{\ms{C}})U^{\dagger}(\tau)]$. Consequently, there is always exists a state $\cat_{\ms{C}}$ which is left unchanged by this channel, hence providing a solution to Eq.~\eqref{Eq:catalytic-constrain}.

One might wonder how often a catalytic evolution leads to a non-classical state of the cavity. In particular, when preparing the cavity in a coherent state $\ket{\alpha}$, does there always exist a state of the catalyst $\cat_{\ms{C}}$ which allows to generate non-classicality? To address this question, we investigate the minimum value of $g^{(2)}$ of the final state $\sigma_{\ms{S}}$, as a function of $|\alpha|$ [see Fig.~~\hyperref[fig:1]{\ref{fig:1}(a)}]. This is done by combining Eqs.~(\ref{eq:g2_conserved}-\ref{eq:corr_term}) and (\ref{Eq-population}-\ref{Eq-coherence}), and imposing a bound on the final time, i.e $g\tau \leq 100$. For $\alpha \in (0, 2]$, we observe that $g^{(2)}(\sigma_{\ms S}) < 1$, indicating a sub-Poissonian statistic for the number of photons and, consequently, evidencing non-classicality. Note that when the initial coherent state has low energy, the final state of the cavity is close to the intial one, but with a slightly reduced variance, leading to a value of $g^{(2)}$ approaching zero.

Although we have focused on the generation of states with Wigner negativity or sub-Poissonian distribution, catalysis also allows the generation of specific non-classical states. In this regard, we show in Appendix~\hyperref[App:generating-squeezed]{C-2} that one can also use our catalytic protocol to generate squeezed states of light. We further determine the possible final states of the cavity that can be obtained under a catalytic protocol.

Let us now ask the converse question, i.e. whether every atomic state can lead to a catalytic evolution. A first observation is that pure states cannot act as useful catalysts in general, and in particular cannot catalytically generate non-classicality. Indeed, a key ingredient for catalysis is the fact that the system and catalyst become correlated [see Eq. (\ref{eq:cat_variance_formula})], which is impossible when the state of the catalyst is pure. 

To further explore this question, we investigated which states of the atom can catalytically generate sub-Poissonian statistics (i.e. $g^{(2)}<1$), given an initial coherent state of the cavity and a limited interaction time $g\tau \leq 100$. In Fig.~\hyperref[fig:1]{\ref{fig:1}(b)}, we display an example of such a set of catalytic states. Interestingly, this set appears to contain states that are almost pure. The structure of the catalytic set is further discussed in Appendix~\hyperref[Appsub:set-catalytic-states]{A-1}, in particular we observe its strong dependence on the intial state of the cavity.

Finally, we investigated scenarios where the cavity's initial state is not a coherent one. Interestingly, we find that catalysis can boost non-classicality. Specifically, starting with an initial  non-classical state $\rho_{\ms{S}}$, one can get a final state $\sigma_{\ms{S}}$ that features more non-classicality, i.e. $g^{(2)}(\sigma_{\ms{S}}) < g^{(2)}(\rho_{\ms{S}})$ (see Appendix ~\hyperref[Appsub:incoherentmixtureg2]{C-2} for details). Conversely, there also exist initial states of the cavity (e.g. Fock states) for which non-classicality cannot be increased catalytically (see Appendix~\hyperref[Appsub:g2Fock]{C-1}). This is intuitive, as Fock states are the most non-classical (i.e. minimize $g^{(2)}$) among all states with a fixed mean energy. As a catalytic protocol must preserve the mean energy, the value of $g^{(2)}$ cannot be increased. 
 
\section{Catalysis in the presence of dissipation}\label{sec:catalysis-realistic-scenarios}

So far we have discussed an idealized (noise-free) scenario in quantum optics to understand the physics behind catalysis in quantum systems. However, any experiment implementing the Jaynes-Cummings model will necessarily feature cavity loss and atomic decay. We will now illustrate that the catalytic effect is robust even after incorporating these two effects in the model. To understand why our results remain valid in the presence of dissipation, it is worth noting that the Perron-Frobenius theorem, which assures that the catalytic constraint can be satisied, is valid for all quantum channels (hence also dissipative). Consequently, even in the presence of losses and noise, the catalytic constraint can still be satisfied. The interesting problem is whether non-classicality can still be generated in such a dissipative protocol. 

In order to understand the role of dissipation in our model let us assume that the cavity $\ms{S}$ is coupled to an environment characterized by temperature $T$ which accounts for the cavity losses. We also consider the possibility of atom decay due to photon emission. Consequently, the dynamics of the atom-cavity system are described by a Lindblad master equation of the form~\cite{Agarwal1986,Agarwal19862,Briegel1993,puri2001mathematical}:
\begin{equation}\label{Eq:rho-diss}
 \dot{\rho}_{\ms{SC}}\!=\!-i\left[H_{\ms{JC}},\rho_{\ms{SC}}\right]+ \kappa(n_{\ms{th}}+1)\mathcal{L}[a] +  \kappa n_{\ms{th}}\mathcal{L}[a^{\dagger}] + \Gamma \mathcal{L}[\sigma_-], 
\end{equation}
where as before we take $\rho_{\ms{SC}} := \dyad{\alpha}_{\ms{S}} \ot \chi_{\ms{C}}$, and $\kappa$ and $\Gamma$ are the cavity and atom dissipation rate, respectively. Moreover, $n_{\ms{th}}=(e^{1/T}-1)^{-1}$ is the average excitation number and \mbox{$\mathcal{L}[L] = L\rho_{\ms{SC}} L^{\dagger} -\frac{1}{2}\{L^{\dagger}L,\rho_{\ms{SC}} \}$} is the Lindblad dissipator with $\{.\,,\,.\}$ denoting the anticommutator. In the case of dissipative evolution \eqref{Eq:rho-diss} the (generalized) catalytic constraint becomes
\begin{equation} \label{eq:gen_catalytic_constraint}
    \sigma_{\ms{C}} := \Tr_{\ms{S}} \mathcal{E}_{\tau}[\rho_{\ms{SC}}] = \chi_{\ms{C}},
\end{equation}
where $\mathcal{E}_{\tau}[\cdot]$ is a quantum channel generated by the evolution from Eq. ~\eqref{Eq:rho-diss} acting for some time $\tau$. 

\begin{figure}[t]
    \centering
    \includegraphics{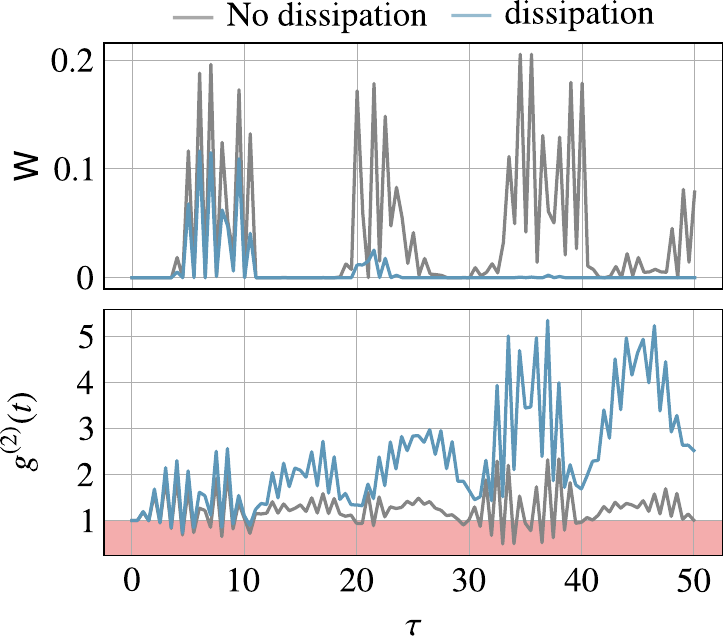}  
    \caption{\label{Fig:catalytic-power}\textbf{Catalysis in realistic scenarios.} Wigner logarithmic negativity and auto-correlation function are plotted as functions of the interaction time $ \tau $. The blue solid line represents the scenario with dissipation, while the gray solid line represents the scenario without dissipation. For each time $ \tau $, the catalytic constraint is satisfied, and the degree of non-classicality activated in the mode is computed. While $\mathsf W \geq 0$ indicates non-classical behaviour, $g^{(2)} < 1$ also witnesses non-classicality (reddish area in the right panel). Both plots use constant dissipation rates: an atom dissipation rate of $ \Gamma = 0.05$ and a cavity dissipation rate of $ \kappa =0.005$. The average number of thermal bath excitations is $ n_{\text{th}} = 0.1 $ and the other parameters are $\omega = 2\pi$, $g = 0.1\pi$ and $\alpha = 1/\sqrt{2}$.} 
\end{figure}

For a given stopping time $\tau$ the solution to Eq.\eqref{Eq:rho-diss} that simultaneously satisfies the catalytic constraint can be determined numerically. In Appendix~\hyperref[App:algorithm]{E} we present an algorithm for constructing the catalytic set for a general dissipative quantum evolution (i.e. any quantum channel), and provide its Python implementation that can be used on a regular laptop.

In order to study the robustness of our results to dissipation we will now analyse two figures of merit, WLN and the auto-correlation function $g^{(2)}$, as a function of the stopping time $\tau$ for closed (reversible) and open (irreversible) quantum dynamics. At all stopping times $\tau$ we ensure that the generalized catalytic constraint from Eq. (\ref{eq:gen_catalytic_constraint}) is met. Our results are summarized in in Fig.~\ref{Fig:catalytic-power} which shows the amount of non-classicality that can be generated in the mode $\ms{S}$ for different choice of interaction times $\tau$. Our first observation is that a catalytic generation of non-classicality is possible even in the presence of dissipation. Secondly, the amount of non-classicality that can be generated in this way generally diminishes with longer interaction times. This is intuitive, as the longer we wait, the higher is the probability of losing photons from the cavity, especially since we are considering a constant dissipation rate for both the cavity and atom. Very interestingly, even for short evolution times, we can achieve relatively high levels of non-classicality as compared to the unitary case. 

Finally, since the catalytic states of the atom $\ms{C}$ are mixed, it must initially be correlated with some external environment. We emphasise that the presence of such correlations does not influence the dynamics of the catalyst $\ms{C}$ and the system $\ms{S}$.

% ------------------------------------------------
% SEC. IV - DISCUSSION
% ------------------------------------------------

\section{Discussion}

We presented a catalytic process for generating non-classical states of light in an optical cavity via interaction with an atom. Our results are valid for any coupling regime in which the Jaynes-Cummings model is applicable and also for realistic scenarios where the presence of noise and loss cannot be neglected. This fundamentally stems from the fact that a catalytic state of the atom exists for any coupling strength and even under the presence of dissipation. This work shows that quantum catalysis, a concept so far explored in the abstract framework of resource theories, is relevant in a practical context (see also Ref.~\cite{YungerHalpern2017}, which discuss the need to turn resource-theory notions into real-world applications). Furthermore, our protocol could potentially be implemented in state-of-the-art experimental setups \cite{frisk2019ultrastrong,periwal2021programmable,wenniger2022coherencepowered}, e.g. in cavity QED \cite{Ballester2012,Blais2021} or trapped ions \cite{Leibfried2003,Lv2018}. Beyond proof-of-principle experiments, it would also be interesting to investigate whether such a catalytic protocol offers a practical advantage. Indeed, the key point of catalysis is that the catalyst is returned exactly in the same state as it was initially prepared. Hence the same atom could in principle be used repeatedly for activating non-classicality in different cavities, or in the same cavity, but at different times. It would be interesting to uncover further instances of quantum catalysis in realistic setups, e.g. exploring other platforms and models in quantum optics, as well as for applications in quantum information and metrology.

Another relevant aspect of our work is that the catalyst (atom) is, in general, a coherent quantum system. Specifically, to generate non-classicality, the atom's state must exhibit coherence in the energy basis over the course of its evolution. This constitutes an example of coherent quantum catalysis~\cite{Lipka_Bartosik_2023}, which contrasts with most previous examples of quantum catalysis. A deeper understanding of the role of coherence in catalysis is an exciting future direction.  

Finally, our analysis also shed light on the key role of correlations in quantum catalysis. This represents a distinctive feature of quantum catalysis. It would be interesting to see if this effect can occur in different physical models. 

\textbf{\textit{Acknowledgements.}} 
We thank Henrik Wilming and Nelly H.Y. Ng for valuable discussions. AOJ acknowledge financial support from the Foundation for Polish Science through the TEAM-NET project (contract no. POIR.04.04.00-00-17C1/18-00); MPL, NB and PLB acknowledge the Swiss National Science Foundation for financial support through the Ambizione grant PZ00P2-186067 and the NCCR SwissMAP.

\bibliography{bibliography}

% ------------------------------------------------
% SEC. II - APPENDIX
% ------------------------------------------------

\appendix
\onecolumngrid

% ------------------------------------------------
% APP. A - Jaynes-Cummings 
% ------------------------------------------------

\section{Jaynes-Cummings model}\label{App:Jaynes-Cummings}
The first part of this appendix provides a brief overview of the eigenproblem solution for the Jaynes-Cummings model (for further information, see references~\cite{Knight2005,Haroche2006}). In the following, we derive the final reduced states of the cavity $\ms{S}$ and the atom $\ms{C}$, and explicitly determine the set of atomic states that satisfy the catalytic constraint [Eq.~\eqref{Eq:catalytic-constrain}]. In what follows, we set $\hbar =1.$

\subsection{Eigenproblem}\label{Appsub:eigenproblem}
The Jaynes-Cummings interaction Hamiltonian $H_{\text{int}} = g \left(a\sigma_+ + a^{\dagger}\sigma_- \right)$ couples pairs of atom-field states $\{\ket{n+1,g},\ket{n,e} \}$. Consequently, the Hamiltonian $H$ decouples into a direct product of $2\times2$-matrix Hamiltonians, i.e.,  $H=\bigoplus_{n=0}^{\infty}H^{(n)}$, where
\begin{equation}\label{Eq:JC_eigeinproblem}
    H^{(n)} \begin{bmatrix}
\ket{n+1,g}\\ \ket{n,e} 
\end{bmatrix} = \begin{bmatrix}
(n+1/2)\omega & g\sqrt{n+1}\, \\ \\
g\sqrt{n+1} &  (n+1/2) \omega 
\end{bmatrix}\begin{bmatrix}
\ket{n+1,g}\\ \ket{n,e} 
\end{bmatrix}.
\end{equation}
The eigenvalue problem for this Hamiltonian yields the eigenfrequencies
\begin{equation}\label{Eq:eigenfrequencies}
    \omega^{(n)}_{\pm} = \qty(n+\frac{1}{2})\omega \pm \frac{1}{2}\mu_n,
\end{equation}
where $\mu_n = 2g \sqrt{n+1} $ is the $n$-photon Rabi frequency. In the resonance regime, the corresponding eigenstates are:
\begin{align}
    \ket{n,+} =   \frac{1}{\sqrt{2}}(\ket{n+1,g} + \ket{n,e}), \qquad 
    \ket{n,-} =  \frac{1}{\sqrt{2}} (\ket{n+1,g} - \ket{n,e}).
\end{align}
The time evolution operator takes the form of
\begin{align}\label{Eq:time-evolution-operator}
    U(t) = &e^{i\omega t/2} \ketbraa{0,g}{0,g}+\sum_{n=0}^{\infty}e^{-i\qty(n+\frac{1}{2})\omega t}  \qty{\cos \frac{\mu_n t}{2}\Bigl(\ketbraa{n+1,g}{n+1,g}+\ketbraa{n,e}{n,e}\Bigl)-i\sin\frac{\mu_n t}{2}\Bigl(\ketbraa{n+1,g}{n,e}+\ketbraa{n,e}{n+1,g}\Bigl)}.
\end{align}

\subsection{Reduced states of subsystems}\label{Appsub:generalsolution}
The unitary operator given by equation~\eqref{Eq:time-evolution-operator} describes the dynamics of the Jaynes-Cummings model in resonance and under the rotating-wave approximation. After the interaction, the joint system \mbox{$\sigma_{\ms SC}(t) = U(t)(\rho_{\ms S} \otimes \cat_{\ms C})U^{\dagger}$} becomes correlated. Let the cavity and the atom be prepared in general mixed states, i.e. \mbox{$\rho_{\ms S} = \sum^{\infty}_{n,m=0}p_{n,m}\ketbraa{n}{m}$} and \mbox{$\cat_{\ms C} = q\ketbraa{g}{g}+ r\ketbraa{g}{e}+r^{*}\ketbraa{e}{g}+(1-q) \ketbraa{e}{e}$}. Then, the reduced state of the cavity at time $t$ is obtained by taking the partial trace over the atom's degrees of freedom, i.e.
$\sigma_\ms{S}(t) := \text{tr}_{\ms C}[{\sigma_{\ms SC}}(t)]$. More specifically,
\begin{align}\label{Eq:cavity-state}
\sigma_{\ms S}(t) = &qp_{0,0} \ketbra{0}{0}+ \sum_{n=0}^{\infty} e^{i(n+1)\omega t}\qty[q p_{0,n+1}\cos\qty(\frac{\mu_n t}{2})+irp_{0,n}\sin\qty(\frac{\mu_n t}{2})]\ketbraa{0}{n+1}+\text{h.c}\nonumber \\&
+\sum_{n,m=0}^{\infty}e^{-i(n-m)\omega t}qp_{n+1,m+1}\qty[\sin\qty(\frac{\mu_n t}{2})\sin\qty(\frac{\mu_m t}{2})\ketbraa{n}{m}+\cos\qty(\frac{\mu_n t}{2})\cos\qty(\frac{\mu_m t}{2})\ketbraa{n+1}{m+1}] \nonumber\\
&+ \sum_{n,m=0}^{\infty}e^{-i(n-m)\omega t}(1-q)p_{n,m}\qty[\cos\qty(\frac{\mu_n t}{2})\cos\qty(\frac{\mu_m t}{2})\ketbraa{n}{m}+\sin\qty(\frac{\mu_n t}{2})\sin\qty(\frac{\mu_m t}{2})\ketbraa{n+1}{m+1}] \nonumber\\
 &+\sum_{n,m=0}^{\infty} i e^{-i(n-m)\omega t}\cos\qty(\frac{\mu_n t}{2}) \sin\qty(\frac{\mu_m t}{2})\qty[p_{n+1,m} r \ketbra{n+1}{m+1}+ p_{n,m+1}r^*\ketbra{n}{m}]+\text{h.c}.
\end{align}
The atomic state is obtained by marginalizing over the photonic degrees of freedom, i.e. $\cat(t) := \text{tr}_{\ms S}[{\sigma_{\ms SC}}(t)]$, which leads to
\begin{align}\label{Eq:catalyst-state-t}
\cat_{\ms C}(t)= q(t)\ketbraa{g}{g}+ r(t)\ketbraa{g}{e}+r^*(t)\ketbraa{e}{g}+[1-q(t)] \ketbraa{e}{e},
\end{align}
Note that we identify $r:=r(0)$ and $q:=q(0)$ in what follows. The coefficients $q(t)$ and $r(t)$ are given by
\begin{align}
 q(t) &= q\sum_{n=0}^{\infty}p_{n}\cos^2\qty(\frac{\mu_{n-1} t}{2})+(1-q)p_n\sin^2\qty(\frac{\mu_n t}{2})+\textrm{Re}[irp_{n+1,n}]\sin{\qty(\mu_n t)} ,\\
r(t) &=-ie^{i\omega t}\sum_{n=0}^{\infty}p_{n,n+1}\sin\qty(\frac{\mu_n t}{2})\cos\qty(\frac{\mu_{n+1}t}{2})+
    re^{i\omega t}\sum_{n=0}^{\infty}p_{n}\cos\qty(\frac{\mu_{n-1} t}{2})\cos\qty(\frac{\mu_{n}t}{2}) +r^{*}\sum_{n=0}^{\infty}p_{n,n+2}\sin\qty(\frac{\mu_n t}{2})\sin\qty(\frac{\mu_{n+1}}{2})\nonumber\\
    &\hspace{7.9cm}+ ie^{i\omega t}q\sum_{n=0}^{\infty}p_{n,n+1}\sin\qty(\frac{\mu_{n}t}{2})\qty[\cos\qty(\frac{\mu_{n-1} t}{2})+\cos\qty(\frac{\mu_{n+1}t}{2})].
\end{align}

\subsection{Catalytic constraint}\label{Appsub:set-catalytic-states}
Here we determine the set of atomic states that evolve catalytically by explicitly solving the catalytic constraint from Eq.~(\ref{Eq:catalytic-constrain}). More specifically, we are looking for the solution to the following operator equation:
\begin{align}\label{Eq:operator-equation}
    \cat_{\ms C}(\tau) = \Tr_{\ms{S}}\{U(\tau)[\rho_{\ms S} \otimes \cat_{\ms C}(\tau)] U(\tau)^{\dagger}\},
\end{align} 
for a fixed time $\tau$. From this point forward, we will abbreviate the diagonal elements of the state $\rho_{\ms{S}}$ as $ p_n := p_{n,n}$. To obtain the set of states that satisfy Eq.\eqref{Eq:operator-equation}, we first define the auxiliary functions:
\begin{align}\label{Eq:auxiliary-equation}
\begin{split}
    \tilde{a}_1(t) &= e^{i\omega t}\sum_{n=0}^{\infty} p_n\cos(gt\sqrt{n})\cos(gt\sqrt{n+1}) -1, \\
    \tilde{a}_2(t) &= ie^{i \omega t}\sum_{n=0}^{\infty}p_{n,n+1}\sin(gt\sqrt{n+1})\qty[\cos(gt\sqrt{n})+\cos(gt\sqrt{n+2})],  \\
    \tilde{a}_3(t) &= e^{i\omega t}\sum_{n=0}^{\infty} p_{n,n+2} \sin(gt\sqrt{n+1})\sin(gt\sqrt{n+2}), \\
    \tilde{a}_4(t) &= -ie^{i\omega t}\sum_{n=0}^{\infty}p_{n,n+1}\sin(gt\sqrt{n+1})\cos(gt\sqrt{n+2}). 
\end{split}
\end{align}
Note that $\tilde{a}_i = e^{i\omega t}a_i$. Next, we observe that Eq.~\eqref{Eq:operator-equation} gives rise to a set of two equations with two variables. By considering the ground state occupation $q(t)$, we find that the states satisfying Eq.~\eqref{Eq:operator-equation} are given by:
\begin{equation}\label{Eq:population_rawq}
    q(t) = \frac{\sum_{n=0}^{\infty}p_n\sin^2(gt\sqrt{n+1})+\operatorname{Re}\qty[ir(t)p_{n+1,n}]\sin(2gt\sqrt{n+1})}{\sum_{n=0}^{\infty}(p_n+p_{n+1})\sin^2(gt\sqrt{n+1})},
\end{equation}
whereas the coherence $r(t)$ obeys the equation 
\begin{equation}\label{Eq:p_and_q_relation}
    r(t) \tilde{a}_1(t) + q(t) \tilde{a}_2(t) + r^*(t) \tilde{a}_3(t) + \tilde{a}_4(t) = 0,
\end{equation}
whose solution is given by
\begin{equation}\label{Eq:r_solutionr}
    r(t) = \frac{\tilde{a}_3(t) \tilde{a}^{*}_4(t)-\tilde{a}^{*}_1(t)\tilde{a}_4(t)}{|\tilde{a}_1(t)|^2 - |\tilde{a}_3(t)|^2}+\frac{\tilde{a}_3(t) \tilde{a}^{*}_2(t)-\tilde{a}^{*}_1(t)\tilde{a}_2(t)}{|\tilde{a}_1(t)|^2 - |\tilde{a}_3(t)|^2}q(t).
\end{equation}
Substituting Eq.~\eqref{Eq:r_solutionr} into Eq.~\eqref{Eq:population_rawq}, we find that
\begin{equation}\label{Eq:q_solution}
    q(t) = \frac{\sum_{n=0}^{\infty}p_n\sin^2(gt\sqrt{n+1})+\operatorname{Re}\qty[i\qty(\frac{\tilde{a}_3(t) \tilde{a}^{*}_4(t)-\tilde{a}^{*}_1(t)\tilde{a}_4(t)}{|\tilde{a}_1(t)|^2 - |\tilde{a}_3(t)|^2})p_{n+1,n}]\sin(2gt\sqrt{n+1})}{\sum_{n=0}^{\infty}(p_{n+1}+p_n)\sin^2(gt\sqrt{n+1})]-\operatorname{Re}\qty[i\qty(\frac{\tilde{a}_3(t) \tilde{a}^{*}_2(t)-\tilde{a}^{*}_1(t)\tilde{a}_2(t)}{|\tilde{a}_1(t)|^2 - |\tilde{a}_3(t)|^2})p_{n+1,n}]\sin(2gt\sqrt{n+1})}.
\end{equation}
Therefore, for a given value of $g$ and time $\tau$, Eqs.~\eqref{Eq:r_solutionr}~and~\eqref{Eq:q_solution} uniquely determine a state of the catalyst. 

% ------------------------------------------------
% APP. B - Catalytic mechanism
% ------------------------------------------------

\section{Quantum catalysis and higher moments of observables}\label{App:higher-moments-of-observables}

In this Appendix we extend the analysis of a catalytic evolution presented in the main text. In particular, we derive an expression for the $k$-th moment of an arbitrary observable $O_{\ms{S}}$ on the system $\ms{S}$ undergoing a catalytic evolution. Specifying this expression to the second moment $(k = 2)$  and the particle number observable $(O_{\ms{S}} = n_{\ms{S}})$ leads to Eq. (\ref{eq:cat_variance_formula}) as stated in the main text. Then, using the results of Appendix \ref{App:Jaynes-Cummings}, we obtain an explicit expression for the second moment in a catalytic evolution as specified by the Jaynes-Cummings Hamiltonian. 

\subsection{Higher moments of observables under catalytic evolution}\label{Appsub:highermomentsofobservables} 

Let $O_{\ms{SC}} = O_{\ms{S}}\ot \mathbbm{1}_{\ms{C}} + \mathbbm{1}_{\ms{S}} + O_{\ms{C}}$ be and additive and conserved observable on the joint system $\ms{SC}$. In a catalytic protocol governed by a unitary $U$ we have $\sigma_{\ms{SC}} = U(\rho_{\ms{S}} \ot \cat_{\ms{C}})U^{\dagger}$ such that $\sigma_{\ms{C}} = \cat_{\ms{C}}$. The conservation assumption means that $[U, O_{\ms{SC}}] = 0$. The $k$-th moment of observable $O$ on the system $\ms{S}$ before and after the catalytic process are respectively given by
\begin{align}
    \langle O^k_{\ms{S}} \rangle_{\rho} := \Tr[O_{\ms{S}}^k \rho_{\ms{S}}], \qquad \langle O^k_{\ms{S}} \rangle_{\sigma} := \Tr[O_{\ms{S}}^k \sigma_{\ms{S}}] = \Tr[(O_{\ms{S}}^k \ot \mathbbm{1}_{\ms{C}}) \sigma_{\ms{SC}}].
\end{align}
Let us write $O_{\ms{SC}}^k = O_{\ms{S}}^k\ot \mathbbm{1}_{\ms{C}} + \mathbbm{1}_{\ms{C}} \ot O_{\ms{C}}^k + \Delta_k$, where we defined an auxiliary observable
\begin{align}
    \Delta_k := O_{\ms{SC}}^k - O_{\ms{S}}^k\ot \mathbbm{1}_{\ms{C}} - \mathbbm{1}_{\ms{C}} \ot O_{\ms{C}}^k = \sum_{i=1}^{k-1} \binom{k}{i} O^{k-i}_{\ms{S}} \ot O^i_{\ms{C}}.
\end{align}
For example, for $k=2$ we have $\Delta_2 = 2 O_{\ms{S}} \ot O_{\ms{C}}$. Using this to express $\langle O^k_{\ms{S}} \rangle_{\sigma}$ yields
\begin{align}
    \label{eq:moments1}
    \langle O^k_{\ms{S}} \rangle_{\sigma} = \Tr[O_{\ms{SC}}^k\sigma_{\ms{SC}}] - \Tr[(\mathbbm{1}_{\ms{S}} \ot O_{\ms{C}}^k)\sigma_{\ms{SC}}] - \Tr[\Delta_k \sigma_{\ms{SC}}] = \Tr[O_{\ms{SC}}^k\sigma_{\ms{SC}}] - \langle O_{\ms{C}}^k\rangle_{\sigma} - \Tr[\Delta_k \sigma_{\ms{SC}}].
\end{align}
Importantly, due to the conservation law $[U, O_{\ms{SC}}] = 0$, we have $\Tr[O_{\ms{SC}}^k \sigma_{\ms{SC}}] = \Tr[O_{\ms{SC}}^k \rho_{\ms{SC}}] = \langle O_{\ms{S}}^k\rangle_{\rho} + \langle O_{\ms{C}}^k\rangle_{\rho} + \Tr[\Delta_{k} {\rho_{\ms{SC}}}]$. Therefore we can rewrite Eq. (\ref{eq:moments1}) as
\begin{align}
     \langle O^k_{\ms{S}} \rangle_{\sigma} =  \langle O_{\ms{S}}^k\rangle_{\rho} + \langle O_{\ms{C}}^k\rangle_{\rho} - \langle O_{\ms{C}}^k\rangle_{\sigma} + \Tr[\Delta_k (\rho_{\ms{SC}}-\sigma_{\ms{SC}})]  = \langle O_{\ms{S}}^k\rangle_{\rho}  + \Tr[\Delta_k (\rho_{\ms{SC}}-\sigma_{\ms{SC}})], 
\end{align}
where we used the fact that $\langle O_{\ms{C}}^k\rangle_{\rho} = \langle O_{\ms{C}}^k\rangle_{\sigma}$ due to the catalytic constraint. By taking $k=2$ we obtain the desired result.

\subsection{Second moment of photon statistics in the catalytic Jaynes-Cummings evolution}\label{Appsub:second-moment-photon}
Consider the second moment of photon statistics
\begin{align}
    \langle n^2_{\ms{S}} \rangle_{\sigma} = \langle n^2_{\ms{S}} \rangle_{\rho} + 2\left[ (1-q)\,\langle n_{\ms{S}} \rangle_{\rho}  - \langle n_{\ms{S}} \ot \dyad{e}_{\ms{C}}\rangle_{\sigma}\right],
\end{align}
where $q := \langle e |\cat_{\ms{C}}|e \rangle$ is the excited-state occupation of the catalyst,  $\sigma = U(\rho_{\ms{S}} \ot \cat_{\ms{C}})U^{\dagger}$ and $\Tr_{\ms{S}}[\sigma] = \cat_{\ms{C}}$. Let us focus on the following term:
\begin{align}\label{Eq:term-second-moment}
     \langle n_{\ms{S}} \ot \dyad{e}_{\ms{C}}\rangle_{\sigma} &= \Tr\left[U^{\dagger}(n_{\ms{S}} \ot \dyad{e}_{\ms{C}})U (\rho_{\ms{S}} \ot \cat_{\ms{C}})\right] = \sum_{k=0}^{\infty} k \Tr[U^{\dagger}\dyad{k, e} U (\rho_{\ms{S}} \ot \cat_{})],  
\end{align}
Using Eq.~(\ref{Eq:time-evolution-operator}), we can write
\begin{align}
    U^{\dagger} \ket{k, e} &= e^{i(k+\frac{1}{2})\cat t} \left(c_k \ket{k,e} - i s_k \ket{k+1, g}\right),
\end{align}
where $c_k := \cos(g t \sqrt{k+1})$ and $s_k := \sin(g t \sqrt{k+1})$. Substituting the above result into Eq.~\eqref{Eq:term-second-moment} leads to
\begin{align}
    \langle n_{\ms{S}} \ot \dyad{e}_{\ms{C}}\rangle_{\sigma} = \sum_{k = 0}^{\infty} k \left[(1-q)c_k^2 p_{k,k} + 2 s_k c_k \text{Im}(p_{k+1,k} r) + (1-q) s_{k}^2 p_{k+1,k+1} \right].
\end{align}

% ------------------------------------------------------------------------------
% APP. C - Exploring catalytic evolutuin in the Jaynes-Cummings model
% ------------------------------------------------------------------------------

\section{Exploring catalytic evolution in the Jaynes-Cummings model}\label{App:set-catalytic-states}

In this Appendix we examine different aspects of the catalytic evolution in the Jaynes-Cummings model. 

\subsection{Which states of the cavity cannot have their non-classicality enhanced?}\label{Appsub:g2Fock}

Let us start by examining Eqs.~\eqref{Eq:r_solutionr}~and~\eqref{Eq:q_solution} under the assumption that the initial state of the cavity is an incoherent mixture of Fock states, i.e. $\rho_{\ms{S}} = \sum_{n,m} p_{n,m}\dyad{n}{m}$ with $p_{n,m} = 0$ if $n \neq m$.  In this case, from Eqs. (\ref{Eq-population}) and (\ref{Eq-coherence}) we can infer that the only feasible states of the catalyst are those with $q_{\text{coh}} = 0$ and $r = 0$. Consequently, the atomic state is incoherent in the energy basis and its ground state occupation takes the form of
\begin{equation}\label{Eq:incoherent-catalyst}
    q = \frac{\sum_{n=0}^{\infty}p_n\sin^2(gt\sqrt{n+1})}{\sum_{n=0}^{\infty}(p_n+p_{n+1})\sin^2(gt\sqrt{n+1})} .
\end{equation}

Let us now demonstrate that the second-order auto-correlation function $g^{(2)}$ of a pure Fock state, i.e. $\rho_{\ms{S}} = \dyad{k}_{\ms{S}}$, cannot decrease under a catalytic evolution. To do this, we will prove that the quantity
\begin{equation}
    g^{(2)}(\sigma_{\ms S}) - g^{(2)}(\rho_{\ms S}) = \frac{\Tr[(a^{\dagger 2}a^2)\sigma_{\ms S}]}{\Tr[(a^{\dagger} a) \sigma_{\ms S}]} - \frac{\Tr[(a^{\dagger 2}a^2)\rho_{\ms S}]}{\Tr[(a^{\dagger} a) \rho_{\ms S}]} = \frac{\Tr[(a^{\dagger 2}a^2)(\sigma_{\ms S}-\rho_{\ms S})]}{\Tr[(a^{\dagger} a) \rho_{\ms S}]}
\end{equation}
is nonnegative. This will be accomplished by showing that the following inequality holds: 
\begin{equation}\label{Eq:inequality-g}
 \Tr[(a^{\dagger 2}a^2)(\sigma_{\ms S}-\rho_{\ms S})] \geq 0 .   
\end{equation}
In what follows, we will omit the index $\ms{S}$ as well as any explicit reference to the variables' dependence on time. When the initial state of the cavity is a Fock state, then Eq.~\eqref{Eq:cavity-state} implies that the state of the cavity after the catalytic evolution is given by
\begin{align}\label{Eq:cavity-state-fock2}
\!\!\!\sigma = \qty[(1-q)\cos^2\qty(gt\sqrt{k+1})+q\cos^2\qty(gt\sqrt{k})]\ketbra{k}{k}+q\sin^2\qty(gt\sqrt{k})\ketbra{k-1}{k-1}+(1-q)\sin^2\qty(gt\sqrt{k+1})\ketbra{k+1}{k+1},
\end{align}
where $q$ is determined by Eq.~\eqref{Eq:incoherent-catalyst}, which for this particular case takes the form:
\begin{equation}\label{Eq:q_Fock-states}
    q = \frac{\sin^2\qty(gt\sqrt{k+1})}{\sin^2\qty(gt\sqrt{k+1})+\sin^2\qty(gt\sqrt{k})} \quad \text{and} \quad 1-q = \frac{\sin^2\qty(gt\sqrt{k})}{\sin^2\qty(gt\sqrt{k+1})+\sin^2\qty(gt\sqrt{k})}.
\end{equation}
By substituting Eq.\eqref{Eq:q_Fock-states} into Eq.\eqref{Eq:cavity-state} and introducing the notation \mbox{$\psi = \frac{1}{2}(\ketbra{n-1}{n-1}+\ketbra{n+1}{n+1})$}, we obtain:
\begin{equation}
    \sigma = \frac{\sin^2\qty(gt\sqrt{k})\cos^2\qty(gt\sqrt{k+1})+\sin^2\qty(gt\sqrt{k+1})\cos^2\qty(gt\sqrt{k})}{\sin^2\qty(gt\sqrt{k+1})+\sin^2\qty(gt\sqrt{k})}\rho+ \frac{2\sin^2\qty(gt\sqrt{k})\sin^2\qty(gt\sqrt{k+1})}{\sin^2\qty(gt\sqrt{k+1})+\sin^2\qty(gt\sqrt{k})}\psi.
\end{equation}
Alternatively, we can express the above equation as \mbox{$\sigma = t\rho + (1-t)\psi$}, where
\begin{equation}
t = \frac{\sin^2\qty(gt\sqrt{k})\cos^2\qty(gt\sqrt{k+1})+\sin^2\qty(gt\sqrt{k+1})\cos^2\qty(gt\sqrt{k})}{\sin^2\qty(gt\sqrt{k+1})+\sin^2\qty(gt\sqrt{k})}.
\end{equation}
With these results at hand, we can manipulate Eq.~\eqref{Eq:inequality-g} to obtain
\begin{align}
    \Tr[(a^{\dagger 2}a^2)(\sigma-\rho)] &= (1-t)\Tr[(a^{\dagger 2}a^2)(\psi-\rho)] = \frac{(1-t)}{2}\Tr[(a^{\dagger 2}a^2)(\ketbra{k+1}{k+1}+\ketbra{k-1}{k-1})]-(1-t)\Tr[(a^{\dagger 2}a^2)\ketbra{k}{k}] \nonumber \\
    &= \frac{(1-t)}{2}\qty[(k+1)k + (k-1)(k-2) - 2k(k-1)] = (1-t) \geq 0.    
\end{align}
Thus, we conclude that second-order coherence $g^{(2)}$ can only increase during a catalytic processes involving a pure Fock state, i.e.
\begin{equation}
    g^{(2)}(\sigma) - g^{(2)}(\rho) \geq 0.
\end{equation}
It is not clear whether there exist other states of the cavity whose non-classicality, as witnessed by the $g^{(2)}$ function, cannot be further enahnced. We leave this interesting problem for future work.

\subsection{Generating squeezed states}\label{App:generating-squeezed}

The state of the cavity after the catalytic process is described by Eq.~\eqref{Eq:cavity-state}, with $q$ and $p$ given by Eqs.~\eqref{Eq:q_solution} and~\eqref{Eq:r_solutionr}. In the main text, we have discussed that both states with Wigner negativity and sub-Poissonian states can be produced catalytically, but the specific form of the states of light was not directly addressed. Here, we focus on this question and show that it is possible to prepare squeezed states in a catalytic way.

Defining the field quadratures by $X_1 = (a^{\dagger}+a)/\sqrt{2}$ and $X_2 = i(a^{\dagger}-a)/\sqrt{2}$, which satisfy the commutation relation $[X_1,X_2]=i$. The Heisenberg uncertainty principle is given by $\Delta X_1 \Delta X_2 \geq 1/2$, where $(\Delta X_1)^2 = \langle X_1^2\rangle - \langle X_1\rangle^2$ and similarly for $(\Delta X_2)^2$. For the coherent field state, the uncertainties are equal to $\Delta X^{(\alpha)}_1 = \Delta X^{(\alpha)}_2 = 1/\sqrt{2}$. A field state is called squeezed if the uncertainty of one of the quadratures is below the vacuum level, i.e., $\Delta X_1 < 1/\sqrt{2}$ or $\Delta X_2 < 1/\sqrt{2}$. Here, we consider the squeezing parameter
\begin{equation}
    \xi = \frac{\Delta X_1}{\Delta X^{(\alpha)}_1} = \sqrt{2} \Delta X_1
\end{equation}
and search for conditions where $\xi < 1$, which manifests field squeezing.

\begin{figure*}
    \centering
    \includegraphics{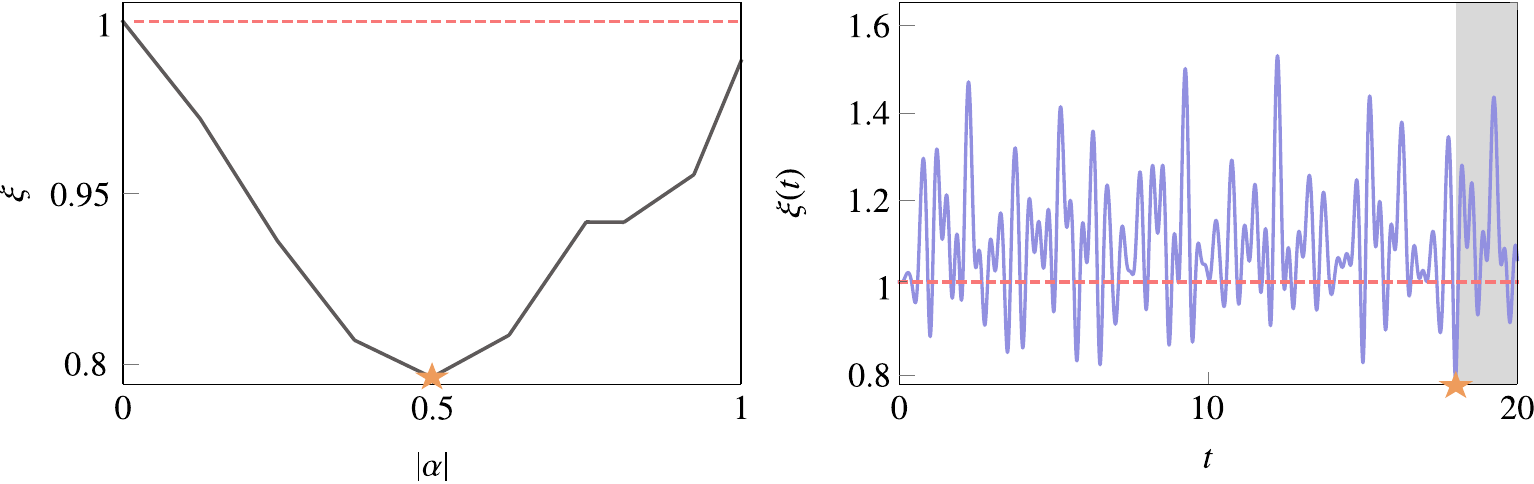}  
    \caption{\textbf{Generating squeezed states via quantum catalysis.} The left panel shows the minimal value of $\xi$ obtained as a function of the amplitude $|\alpha|$ of the initial coherent state of the cavity within the interval of interaction time $\tau \leq [0,100]$. Taking the minimal value of $|\alpha|$, the right panel displays its time evolution, defined as $\xi(t) := \xi (\sigma_{\ms{S}}(t))$. The orange star indicates the final time ($\tau \approx 18$) at which the evolution becomes catalytic. At this point, the atom returns to its initial state, while a squeezed state with $\xi(\tau) \approx 0.79 < \xi(0) = 1$ has been produced. The values of the parameters are: $\alpha = 1/2$, $\omega = \pi$, and $g = 2\pi$.  \label{F:squeezing-fig} } 
\end{figure*}

The catalytic generation of squeezed states can be observed in Fig~\ref{F:squeezing-fig}. The left plot depicts the minimal value of $\xi$ obtained as a function of the amplitude $|\alpha|$ of the initial coherent state of the cavity. We impose a limit on the interaction time $\tau \leq 100$ and observe that $\xi$ goes below the shot noise level, implying the presence of squeezing. In the right plot, we depict the evolution of $\xi$. Catalysis occurs at time $\tau = 18$ (represented by an orange star). Starting from the initial state of the cavity $\rho_{\ms S} = \ketbra{\alpha}_{\ms S}$, which has $\xi = 1$, we obtain a final state $\sigma_{\ms S}$ for which $\xi \approx 0.79$.

\subsection{Which states of the atom allow to generate non-classicality?}

An interesting question is which states of the atom can satisfy the catalytic constraint from Eq. (\ref{Eq:catalytic-constrain}) for some value of the stopping time $\tau$, and for a fixed initial (coherent) state of the cavity, i.e. $\rho_{\ms{S}} = \dyad{\alpha}$ with amplitude $|\alpha|^2$. In what follows we will refer to the set of all atomic states that satisfy Eq. (\ref{Eq:catalytic-constrain}) as the \emph{set of catalytic states}. Moreover, we observe that not every state in this set can lead to a final state of the cavity that is non-classical.

\begin{figure*}
    \centering
    \includegraphics{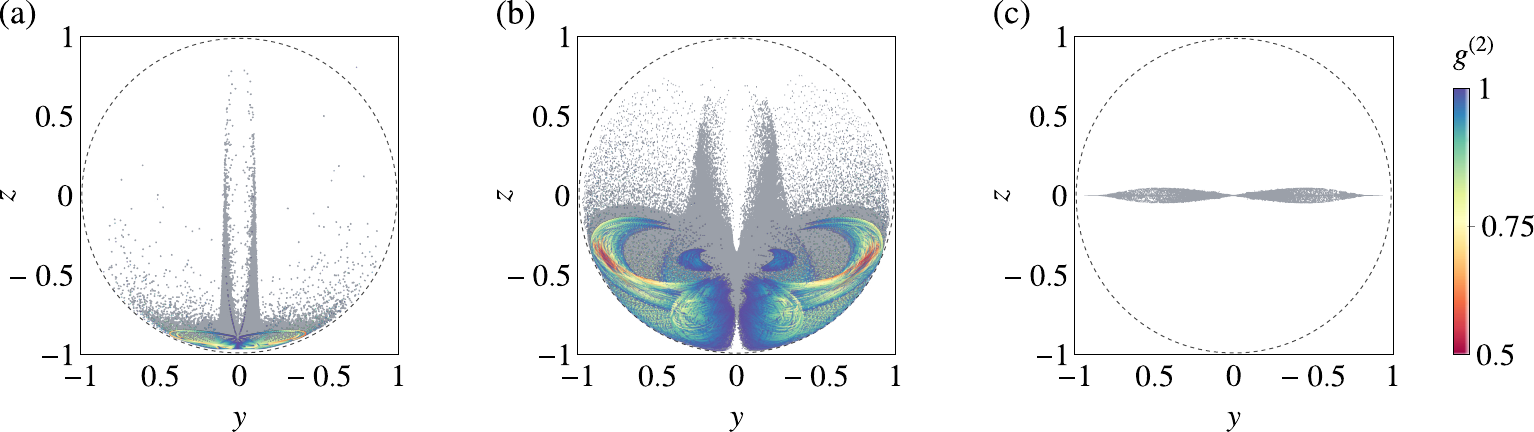}  
    \caption{\label{Fig:set-of-catalytic-states}\textbf{Set of catalytic state for different initial state preparation}. Atomic states (projection of the $y-z$ plane of the Bloch ball) that satisfy the catalytic constraint (gray) for (a) $\alpha = 0.2$, (b) $\alpha = 1/\sqrt{2}$, and (c) $\alpha = 25$ are highlighted by a uniform gradient if they produce $g^2(\sigma_\ms{S}) < 1$ All panels were generated by imposing a limit on interaction time with $\tau \leq 100$ and taking $10^{6}$ samples. The parameters are $\omega = 2\pi$ and $g = \pi$.
    } 
\end{figure*}

To investigate these questions, we characterize numerically the catalytic set for three initial coherent states of the cavity corresponding to $\alpha \in \{0.2, 1/\sqrt{2}, 25\}$. Results are presented in Fig.~\ref{Fig:set-of-catalytic-states}, where the catalytic set is shown in grey. Additionally, we also determine which states in the catalytic set can generate non-classicality, focusing on the $g^{(2)}$ function here. Similarly to Fig.~\hyperref[fig:1]{\ref{fig:1}(b)} of the main text, we represent these states in colour, the latter indicating the level of non-classicality being generated. Interestingly, these sets vary significantly with $|\alpha|$. Also, when $|\alpha|$ is large, catalytic states are distributed close to the equatorial plane of the Bloch sphere, and we could find no instance where non-classicality is generated. 

\subsection{Boosting non-classicality via a catalytic process for incoherent mixtures of Fock states}\label{Appsub:incoherentmixtureg2}
For an incoherent mixture of Fock states \mbox{$\rho  = \sum_{n} p_n \ketbra{n}{n}$}, the second-order coherence is given by
\begin{equation}\label{Eq-app-second-order}
    g^{(2)}(\rho) = \frac{\sum_{n=0}^{\infty}n(n-1) p_n}{(\sum_{n=0}^{\infty}n p_n)^2}.
\end{equation}
Assuming that the initial state of the cavity is prepared in a state $\rho_{\ms S} = \frac{1}{4}\ketbra{0}{0} +\frac{3}{4}\ketbra{2}{2})$, then its second-order coherence is $g^{(2)}(\rho_{\ms S}) = 2/3$. According to Eq.~\eqref{Eq:cavity-state}, the state of the cavity after the catalytic protocol (at time $t = \tau$) takes the form of 
\begin{align}
    \sigma_{\ms S} = \frac{1}{4}\qty[q+(1-q)\cos^2 g t] \ketbra{0}{0}+\frac{3q}{4}\qty[\sin^2 \qty(g\tau\sqrt{2})+\frac{(1-q)}{4}\sin^2 g t]\ketbra{1}{1}&+\frac{3}{4}\qty[q\cos^2 \qty(g\tau\sqrt{2})+(1-q)\cos^2 \qty(g\tau\sqrt{3})]\ketbra{2}{2}\nonumber\\&+\frac{3(1-q)}{4}\sin^2 \qty(g\tau\sqrt{3})\ketbra{3}{3}
\end{align}
where $q$ is determined by Eq.~\eqref{Eq:incoherent-catalyst}:
\begin{equation}
    q = \frac{\sin^2\qty(g\tau)+3\sin^2\qty(g\tau\sqrt{3})}{\sin^2\qty(g\tau)+3\qty[\sin^2\qty(g\tau\sqrt{3})+\sin^2\qty(g\tau\sqrt{2})]}.
\end{equation}
Using Eq.~\eqref{Eq-app-second-order}, the second-order coherence for the final state is given by
\begin{equation}
 g^{(2)}(\sigma_{\ms S}) = \frac{2}{3}\qty{q\cos^2 \qty(g\tau\sqrt{2})+(1-q)\qty[1+2\sin^2 \qty(g\tau\sqrt{3})]}.
\end{equation}
Therefore, for $g\tau = 7.5\pi$, we obtain $g^{(2)} \approx 0.505$, indicating that non-classicality in the mode was catalytically increased. 

Interestingly, when the cavity is initially prepared in a mixture of Fock states, we oberve that the catalyst must necessarily be in an incoherent state for satisfying the catalytic constraint.

\section{Algorithm for the evolution of dissipative quantum systems under the catalytic constraint}\label{App:algorithm}

We now present an algorithm devised to uncover the catalytic set for a dissipative quantum system governed by a given Lindbladian superoperator $\mathcal{L}$. The algorithm harnesses the power of semidefinite programming (SDP) to compute the state of the catalyst $\chi_{\ms C}$ for a given specified time.

\begin{pabox}[label={algorithmcounter}]{Dissipative catalyst evolution finder}
	\begin{enumerate}[leftmargin=0.2cm]

		\item \textbf{Setup.}
		\begin{enumerate}[leftmargin=0.2cm]
            \item Input the initial state $\rho_{\ms S}$, the Liouvillian superoperator $\mathcal{L}$, and a list `\texttt{times}' with the desired time points.
			\item Initialize empty lists \texttt{states}, \texttt{catalysts}, and \texttt{Kraus} for storing results.
		\end{enumerate}

		\item \textbf{Time Evolution.}
		\begin{enumerate}[leftmargin=0.2cm]
            \item For each time $t$ in \texttt{times}, compute the evolution operator $\mathcal{L}_t = e^{Lt}$. 
            \item Convert $\mathcal{L}_t$ into its corresponding Kraus operators $K_i$ and add them to the \texttt{Kraus} list.
			\item Set up a semidefinite program with the variable $\chi_{\ms C}$, subject to the following constraints:
                \begin{enumerate}[label=\alph*.]
					\item $\chi_{\ms C}$ is positive semidefinite: $\chi_{\ms C} \geq 0$.
					\item $\chi_{\ms C}$ has a trace of one: $\tr[\chi_{\ms C}] = 1$.
					\item $\chi_{\ms C}$ fulfills the catalytic constraint: $\chi_{\ms C} = \tr_{\ms S}[\sum_{i}K_i (\rho_{\ms S}\otimes \chi_{\ms C})K^{\dagger}_i]$.
				\end{enumerate}
            \item Solve the semidefinite program and append the resulting $\chi_{\ms C}$ state to the \texttt{catalysts} list.
			\item Calculate the joint state $\sigma_{\ms{SC}} = \sum_i K_i (\rho_{\ms S} \otimes \chi_{\ms C})K^{\dagger}_i$ using the obtained $\chi_{\ms C}$ and add it to the \texttt{states} list.
		\end{enumerate}

		\item \textbf{Return Results.}
		\begin{enumerate}[leftmargin=0.2cm]
			\item Return the \texttt{states} and \texttt{catalysts} lists, which represent the evolution of both the joint system and the catalyst state across the specified time points.
		\end{enumerate}
	\end{enumerate}
\end{pabox}
We provide a \texttt{python} implementation of the aforementioned algorithm, as detailed in~\cite{alexcatalyticonstraint}.

\section{Catalytic protocol in the presence of multiple cavities} \label{app:mcav}

\begin{figure}[t]
    \centering
    \includegraphics{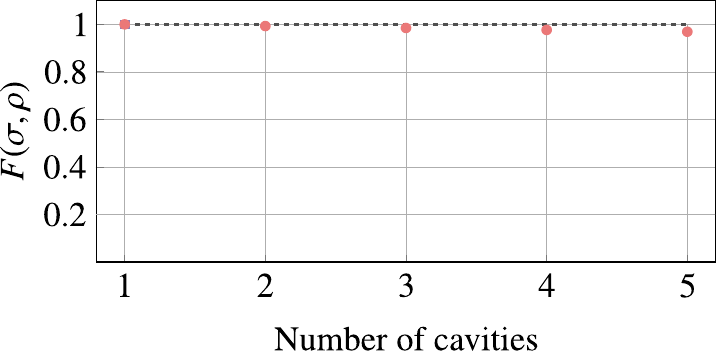}  
    \caption{\label{F-fidelity-multiple-cavities}\textbf{Fidelity as a function of the number of cavities}. In the regime $g\tau \approx \pi $ (red dot), a high fidelity is observed, indicating the feasibility of employing the catalytic protocol to activate the non-classicality across five modes. 
} 
\end{figure}

A relevant question one might have when investigating the catalytic protocol is whether the atom can be reused in a \emph{truly} catalytic way. That is, whether the catalytic protocol can be performed multiple times, while the catalyst is prepared only once. To see whether our protocol fulfills this requirement we consider a single atom $\ms{C}$ interacting sequentially with $N$ cavities $\ms{S} := \ms{S}_1 \ms{S}_2 \ldots \ms{S}_N$ with the same matching mode according to a catalytic process satisfying Eq. (\ref{Eq:catalytic-constrain}). More specifically, we imagine the following protocol:

\begin{enumerate}
\item For a given choice of parameters $\omega$, $g$, and $\tau$, determine the state of the atom $\chi_{\ms C}$ as discussed in Sec. \ref{sec:meachanism}.
\item Let the atom interact sequentially with $N$ cavities with the same frequency $\omega$ and interaction strenght $g$ and for the same amount of time $\tau$. This leads to the following unitary transformation of the composite system
\begin{equation}
    \sigma_{\ms{SC}} = \bigotimes_{i=1}^{N}[U_i\,(\rho_{\ms S_{i}} \otimes \chi_{\ms C})\,U^{\dagger}_i],
\end{equation}
where for all $i \in \{1, ..., N\}$, we have $\rho_{{\ms{S}_i}} = \ketbra{\alpha}{\alpha}_{\ms{S}_i}$ and $U_i$ is the unitary from Eq.~\eqref{Eq:time-evolution-operator} applied to $\ms{S}_i$ and $\ms{C}$.
\item After interacting with all the cavities the atom is removed from the last cavity ($\ms{S}_N$) and allowed to dissipate its energy to the environment. This removes the correlations built between the atom and the cavities. 
\end{enumerate}

After discarding the atom $\ms{C}$ the joint state of the cavities is given by $\sigma_{\ms{S}} = \Tr_{\ms{C}}[\sigma_{\ms{SC}}] = \sigma_{\ms{S}_1 \ms{S}_2 \ldots \ms{S}_n}$. Importantly, each of the states $\sigma_{\ms{S}_i}$ produced in this way is the same and non-classical, and can be further used in some information processing task. 

In certain applications one might be also interested in the amount of correlations developed between the cavities themselves. Such correlations could, in principle, limit the protocol's applicability to certain tasks which require producing multiple uncorrelated non-classical states. Luckily, the amount correlations developed between the cavities after discarding the atom is relatively low. To see this we will use the quantum fidelity $F(\sigma_{\ms{SC}}, \sigma_{\text{target}})$ where $\sigma_{\text{target}}$ is a product state $\sigma_{\text{target}} = \sigma_{\ms{S}_1} \ot \sigma_{\ms{S}_2} \otimes \ldots \otimes \sigma_{\ms{S}_N}$ and $\sigma_{\ms{S}_i} = \sigma_{\ms{S}_j}$ for $1\leq i, j \leq N$.  In Fig.~\ref{F-fidelity-multiple-cavities}, we show the resulting fidelity $F(\sigma_{\ms{SC}}, \sigma_{\text{target}})$ for up to $N = 5$ cavities. Notably, we observe that even for $N = 5$ the fidelity is relatively high, indicating tha the cavities are approximately uncorrelated after the catalytic protocol, and hence can be treated as independent. 

\end{document}